\documentclass[prl,aps,twocolumn,floatfix,superscriptaddress]{revtex4-1}
\usepackage{dcolumn,amsmath}
\usepackage{graphicx}
\usepackage{bm}
\usepackage{hyperref} 
\usepackage{epsfig}
\usepackage{amssymb}

\newcommand{\vect}[1]{\ensuremath{\mathbf{#1}}}
\newcommand{\bra}[1]{\ensuremath{\left\langle #1 \right\vert}}
\newcommand{\ket}[1]{\ensuremath{\left\vert #1 \right\rangle}}

\begin{document}
\title{
Enhanced nuclear spin dependent parity violation effects using the $^{199}$HgH molecule
}
\author{A. J. Geddes}
\affiliation{School of Physics, University of New South Wales, Sydney, New South Wales 2052, Australia}

\author{L. V. Skripnikov}
\affiliation{National Research Centre “Kurchatov Institute” B.P. Konstantinov Petersburg Nuclear Physics Institute, Gatchina, Leningrad district 188300, Russia}
\affiliation{Department of Physics, Saint Petersburg State University, Saint Petersburg, Petrodvoretz 198904, Russia}

\author{A. Borschevsky}
\affiliation{Van Swinderen Institute, University of Groningen, Nijenborgh 4, 9747 AG Groningen, The Netherlands}

\author{J. C. Berengut}
\affiliation{School of Physics, University of New South Wales, Sydney, New South Wales 2052, Australia}

\author{V. V. Flambaum}
\affiliation{School of Physics, University of New South Wales, Sydney, New South Wales 2052, Australia}

\author{T. P. Rakitzis}
\affiliation{Department of Physics, University of Crete, 71003 Heraklion-Crete, Greece}
\affiliation{Institute of Electronic Structure and Laser, Foundation for Research and Technology–Hellas, 71110 Her aklion-Crete, Greece}

\date{13 April 2018}

\begin{abstract}
\noindent
Electron interactions with the nuclear-spin-dependent (NSD) parity non-conserving (PNC) anapole moment are strongly enhanced within heteronuclear diatomic molecules. A novel, low-energy optical rotation experiment is being proposed with the aim of observing NSD PNC interactions in HgH. 
Based on the relativistic coupled cluster method we present a complete calculation of the circular polarization parameter $P = 2\,\textrm{Im}(E1_{PNC})/M1 \approx 3 \times 10^{-6}\; \kappa$ for the $^2\Sigma_{1/2} \to ^2\Pi_{1/2}$ optical transition of HgH, where $\kappa$ is a dimensionless constant determined by the nuclear anapole moment.
This provides an improvement in sensitivity to NSD PNC by 2 -- 3 orders of magnitude over the leading atomic Xe, Hg, Tl, Pb and Bi optical rotation experiments, and shows that the proposed measurement will be sensitive enough to extract the $^{199}$Hg anapole moment and shed light on the underlying theory of hadronic parity violation.

\end{abstract}

\maketitle

\section{Introduction}
The parity operation results in the inversion of the spatial coordinates of the object it acts on. Although many physical systems are symmetric under parity operations, some give rise to different physics under the inversion of spatial coordinates. The violation of symmetry under a parity operation is known as parity non-conservation (PNC). PNC measurements within the $^{133}$Cs atom~\cite{Wood:97}, which are dominated by nuclear-spin-independent (NSI) PNC effects, are in outstanding agreement with predictions from the standard model~\cite{Ginges2002,porsev09prl,Dzuba:2012}. These have placed bounds on the energy at which new physics may be discovered from this process at greater than 0.7~TeV/$c^2$ (see e.g.\cite{Dzuba:2012,Roberts:2014}). Experimental investigation has consequently shifted towards nuclear-spin-dependent (NSD) PNC effects with the aim of testing low energy quantum chromodynamics (QCD) and nuclear theory~\cite{Ginges:2004}.

The nuclear anapole moment is one example of a manifestation of NSD PNC \cite{Khriplovich1980,Sushkov1984} 
and is the main mechanism behind the PNC considered in this letter. Zel'dovich developed the notion of the anapole moment of an elementary particle in 1957 \cite{Zeldovich:57}. Subsequently, Flambaum and Khriplovich proposed the existence of the nuclear anapole moment, which was found to be the dominant NSD PNC effect in heavy atoms and molecules 
\cite{Sushkov:78,Khriplovich1980}. 

The observable NSD PNC effects of the nuclear anapole moment include manifestations of the parity violating electric dipole transition (E1$_{PNC}$) in atoms and molecules. PNC effects have small amplitudes compared to molecular and atomic electromagnetic processes and are difficult to detect \cite{Kozlov:95}. 
The nuclear anapole moment has been detected only once within the $^{133}$Cs atom \cite{Wood:97} (where NSD PNC is sub-dominant) as experimental techniques have lacked the sensitivity to detect NSD PNC effects with certainty. NSD PNC calculations in molecules provides a new window of opportunity to study parity violating nuclear forces which create the nuclear anapole moment.

PNC effects are enhanced within diatomic molecules due to closely spaced rotational levels of opposite parity \cite{Sushkov:78,Labzowsky:78}. In this Letter we show that
mercury hydride (HgH) in particular is a promising choice for the study of PNC effects, not only because it gives an enhanced, pure NSD PNC signal but also because it is easy to make at room temperature. These effects manifest as $E1_{PNC}$ transitions that violate the parity selection rules of dipole transitions. The transition can be detected via interference of the $E1_{PNC}$ amplitude with an allowed $M1$ transition amplitude between the same states. This results in the rotation of the polarisation plane of light passing through a gas of HgH molecules, which is referred to as PNC optical rotation \cite{Khriplovich:91}
\begin{equation}
\phi_{PNC} = -\frac{4 \pi l}{\lambda}(n(\omega) - 1) \frac{\textrm{Im}(E1_{PNC})}{M1}
\label{eq:rota}
\end{equation}
which depends on the experimental parameters $\omega$, $n(\omega)$, $l$ and $\lambda$ which are the optical frequency, refractive index due to the absorption line, the path length of light, and the optical wavelength, respectively.

The experimental techniques developed in~\cite{Bougas:2012} promise greater sensitivity in NSD PNC measurements of this type. The experimental set-up includes a cavity in which four mirrors are placed in a ``bow-tie'' configuration, allowing polarised light to make multiple passes through the cavity before detection. 
By increasing the path length of light passing through the sample within the optical cavity, the experiment is expected to enhance optical rotation signals by up to 4 orders of magnitude.

For small optical paths $l<2L$, where $L$ is absorption length at a given frequency off-resonance, the optical rotation $\phi_{PNC}$ increases linearly with the sample density (or the number of cavity passes); it reaches $\phi_{PNC} \sim P = 2 \frac{\textrm{Im}(E1_{PNC})}{M1}$ at  $\omega-\omega_r =\Delta_D$ when $l=2L$ (where the maximum signal-to-noise ratio is achieved), i.e. at transmission $1/e^2 = 13.5\%$. Here, $\omega_r$ is the resonant frequency and $\Delta_D$ is the Doppler width which is much larger than the natural width. However for $l>2L$ in the resonance, larger values of $\phi_{PNC}$ can be found by tuning the wavelength further off resonance: absorption falls rapidly as $1/(\omega - \omega_r)^2$ while $\phi_{PNC}$ falls slower as $1/(\omega - \omega_r)$. Therefore to suppress absorption, one must go to the tail of the resonance which will result in large $L$ and $\phi_{PNC}$ much larger than $P$. In order to achieve this we must have sufficiently large effective $l$ after many reflections of light in the cavity \cite{Khriplovich:91}. 




The advantage of the HgH molecule for the PNC experiment is the large rotational constant, which allows   optical transitions to be resolvable for levels of opposite parity.
In this Letter we have performed relativistic coupled cluster calculations of the weak interaction (anapole) matrix elements in the $A_1~^{2}\Pi_{\frac{1}{2}}$ excited state and the ground $X~^{2}\Sigma$ state of $^{199}$HgH, as well as the corresponding E1 and M1 transition amplitudes. These calculations allow for a complete extraction of the nuclear anapole moment of $^{199}$Hg from the proposed experiment.

\section{Spin-rotational Hamiltonian}
$^{199}$HgH is a heteronuclear diatomic molecule with one valence electron. The total valence electronic angular momentum can be expressed as $\vect{J}_e = \vect{S} + \vect{L}$ where \vect{S} is the electron spin and \vect{L} is the orbital angular momentum. HgH has electronic ground state of $X~^2\Sigma_{1/2}$ and first electronic excited state $A_1~^2\Pi_{1/2}$. We assign the laboratory frame coordinates \vect{x}, \vect{y} and \vect{z}, in which the molecule rotates with angular momentum \vect{N}. The magnitude of the separation between discrete rotational levels in HgH is governed by the state-specific rotational constant $B$. Rotational angular momenta can couple to the electronic angular momentum to form a vector \vect{J}:
\begin{equation}
\vect{J} = \vect{N} + \vect{J}_e.
\label{eq:J}
\end{equation}
\vect{J} has a projection along the inter-nuclear axis $\Omega$. Furthermore, both $^{199}$Hg and H have nuclear spin, denoted by \vect{I}$_1$ and \vect{I}$_2$ respectively. A general spin-rotational Hamiltonian $H_{sr}$ can be written for both the $X~^2\Sigma_{1/2}$ and $A_1~^2\Pi_{1/2}$ terms \cite{Kozlov:91,Kozlov:95}:
\begin{equation}
\label{SROT}
H_{sr}=B\vect{J}^2+\Delta\vect{J}\cdot\vect{S}^{\prime}+\vect{I}_1\cdot \hat{\vect{A}}_1 \cdot \vect{S}^{\prime} + \vect{I}_2\cdot \hat{\vect{A}}_2 \cdot \vect{S}^{\prime}.
\end{equation}
%
Here $\hat{\textbf{A}}_1$ and $\hat{\textbf{A}}_2$ are second rank axial tensors describing the spin-spin interaction between electrons and the nucleus, and $\Delta$ is the $\Omega$-doubling constant. In the rotating molecular frame described by $\boldsymbol{\xi}$, $\boldsymbol{\eta}$ and $\boldsymbol{\zeta}$, the tensor contractions
\begin{equation}
\label{HFSr}
\vect{I}\cdot\hat{\vect{A}}\cdot \vect{S}^{\prime}=A_{||}\,\vect{I}_{0}\vect{S}_{0}^{\prime}-
A_{\perp}\,\left(\vect{I}_{1}\vect{S}_{-1}^{\prime}+\vect{I}_{-1}\vect{S}_{1}^{\prime}\right),
%
\end{equation}
are determined by the  parallel and perpendicular hyperfine parameters $A_{||}$ and $A_{\perp}$. $\vect{S}^{\prime}$ is the effective spin whose components act on the projection $\Omega$. If we express the tensor components of \vect{S} in the rotating molecular frame we get \cite{Kozlov:87, Dmitriev:92} 
\begin{align}
\vect{S}^{\prime}_{\hat{n}}\ket{\Omega} &= \Omega\ket{\Omega}, \nonumber \\
\vect{S}^{\prime}_{\pm}\ket{\Omega=\mp 1/2} &= \ket{\Omega=\pm 1/2}, \nonumber \\ 
\vect{S}^{\prime}_{\pm}\ket{\Omega=\pm 1/2} &= 0.
\end{align}

%
%
%
The angular momenta coupling scheme in the case of $X~^2\Sigma_{1/2}$ ground state follows that known as Hund's case \textit{b}. The total electronic angular momentum $\vect{J}_e \approx \vect{S}$ since for this state $\Lambda = 0$ where $\Lambda$ is the projection of electronic orbital angular momentum on the molecular axis. Therefore, we can use $\vect{J}_e \approx \vect{S}$ and the substitution $\vect{J}=\vect{N}+\vect{S}$~\cite{Kozlov:91,Kozlov:95}.
Next, the first and second nuclear spin couple in succession \cite{Kozlov:91}:
\begin{gather}
\textbf{F}_1=\textbf{J}+\textbf{I}_1,
\label{eq:couple1} \\
\textbf{F}=\textbf{F}_1+\textbf{I}_2,
\label{eq:couple2}
\end{gather}
Furthermore, the $\Omega$ doubling constant is defined in this scheme to be
$$\Delta=-2B+\gamma,$$
where $\gamma$ is the spin-doubling constant.

Conversely, the $A_1~^2\Pi_{1/2}$ state follows the coupling scheme described by Hund's case \textit{a}. The projection of total angular momentum $\Omega$ in the direction of a unit vector along the internuclear axis $\hat{\mathbf{n}}$ couples to $\mathbf{N}$ to give
\begin{equation*}
\vect{J} = \vect{N} + \Omega\, \hat{\vect{n}}.
\end{equation*}
$\textbf{I}_1$ and $\textbf{I}_2$ couple to $\mathbf{J}$ to form $\textbf{F}_1$ and $\textbf{F}$ in turn, as in Equations~(\ref{eq:couple1}) and (\ref{eq:couple2}).

The basis states that will be used in this work are defined by quantum numbers \ket{J p F_1 F M}, where $p$ is the parity and $M$ is the projection of the total angular momentum $\textbf{F}$ on the lab axis.

\section{Weak Interaction Constants}
The nuclear anapole moment can interact (via its magnetic field) with an electron wavefunction with non-zero total angular momentum \cite{Flambaum:85b}; this is one mechanism behind NSD PNC interactions in atoms and molecules and can be described using a Hamiltonian of the form
\begin{equation}
 H_{P} = \kappa\frac{G_\mathrm{F}}{\sqrt{2}}\,
\boldsymbol{\alpha}\cdot\vect{I}\, \rho(\vect{r}),
\label{HP}
\end{equation}
where \mbox{$G_\mathrm{F}~=~2.22249 \cdot 10^{-14}$~a.u.} is the Fermi coupling constant in atomic units and $\rho (\vect{r})$ is the normalised nuclear density. $\kappa$ is the dimensionless constant determined by nuclear anapole moment to be extracted from experiment. It has been estimated as \cite{Flambaum:85b}
\begin{equation}
\kappa\approx\frac{9}{10} g \left(\frac{\alpha \mu}{m r_{0}} \right) A^{\frac{2}{3}}.
\end{equation}
where $A$ is the number of nucleons in the nucleus, $m$ is the mass of the proton, $\mu$ is the magnetic moment of the external nucleon, $g$ is a dimensionless constant describing the strength of the weak nucleon-nucleus interaction, $\alpha=\frac{1}{137}$ is the fine-structure constant, and $r_{0}=1.2\times 10^{-13}$ cm is the internuclear distance.
 
It is possible to average over fast electron motion to obtain the effective weak interaction coefficient $W_{a}$, which will be constant for a given molecular state. An effective P-odd Hamiltonian can be written as a T-even pseudoscalar formed from the products of the vectors in the system, namely \vect{I}, the effective electron spin $\vect{S}^{\prime}$ and direction of the internuclear axis $\vect{n}$ \cite{Flambaum:85b}. Therefore, in the presence of anapole moment within the Hg nucleus the total Hamiltonian of HgH will also include the following term:
\begin{equation}
H_{\rm eff} =  
W_{a}\kappa\left(\vect{n}\times\vect{S}^{\prime}\right)\cdot \vect{I},
\label{HEFF}
\end{equation}
where $W_{a}$ can be written as
\begin{equation}
W_{a}=\frac{G_\mathrm{F}}{\sqrt{2}} \bra{\Psi_{\Omega=1/2}}
     \rho(\textbf{r}) {\alpha_+}
     \ket{\Psi_{\Omega=-1/2}}.
\label{W_a}
\end{equation} 
$\Psi_{\Omega=1/2}$ can be the $^{2}\Pi_{1/2}$ or $^{2}\Sigma_{1/2}$ state,
$\alpha_+$ is defined as
%
\begin{eqnarray*}
  \alpha_+=\alpha_\xi+\mathrm{i}\alpha_\eta,
\end{eqnarray*}
and $\alpha_\xi, \alpha_\eta$ are the Dirac matrices in the molecular coordinate system. An approximate expression for  $W_{a}$ can be used to check the corresponding calculations and has been found in \cite{Flambaum:85b} to have the form
\begin{equation}
    W_{a} \approx \frac{\epsilon_{s}}{\nu_{s}^{\frac{3}{2}}} \frac{\epsilon_{p}}{\nu_{p}^{\frac{3}{2}}} Ry \frac{2\sqrt{2}}{\sqrt{3} \pi} G_{F} m_e^{2} \alpha^{2} Z^{2} R_{W} \frac{(-1)^{I+\frac{1}{2}-l}(I+\frac{1}{2})}{I(I+1)},
\label{op}	\end{equation}
where the relativistic correction term $R_{W}$ can be written as
\begin{equation}
R_{W}=\frac{2 \gamma + 1}{3} \left(\frac{a_{B}}{2 Z r_{0} A^{\frac{1}{3}}}\right)^{2-2\gamma}.
    \label{eq:rel}
\end{equation}
$\nu_{s}$ and $\nu_{p}$ are the effective quantum numbers for the $s$ and $p$ atomic Hg orbitals respectively, $\epsilon_{s}$ and $\epsilon_{p}$ are weighting coefficients for the contributions of each atomic orbital, $m_e$ is the mass of the electron, $Ry=13.6$ eV is the Rydberg constant, $l$ is the orbital angular momentum of an external unpaired nucleon, $Z$ is the atomic number, and $a_{B}$ is the Bohr radius.

We have calculated the $W_a(^2\Sigma_{1/2})$ and $W_a(^2\Pi_{1/2})$ constants for HgH using two different methods: the first was a Dirac-Hartree-Fock (DHF) calculation performed as a way of checking the scaling relation between $W_a$ and Z; and the second was an accurate coupled-cluster (CC) calculation which we use in our subsequent calculation of the circular polarization parameter $P$.

In the first method, all $W_a(^2\Sigma_{1/2})$ and $W_a(^2\Pi_{1/2})$ constants were calculated with the relativistic program package {\sc dirac15} \cite{DIRAC15} using the DHF method. The DHF method employs the relativistic, multi-electron Dirac Hamiltonian in conjunction with the Hartree-Fock wavefunction. The $W_a$ constants for ZnF and CdH were calculated within the same approach and used to verify that the 
 $W_a$ values scale as expected with the square of the atomic number Z. 
The final values are displayed in the first column of Table~\ref{table:Wa}.

The calculations were carried out at the experimental bond lengths of both the ground and the excited states of the three molecules \cite{NIST}. The heavy Zn, Cd and Hg atoms were described using Dyall's cc-pvqz basis sets \cite{Dyall04,Dyall:07} and for the H atom we used the uncontracted aug-cc-pVTZ basis set \cite{Dunning89}. Finally, we multiplied the output by a core polarisation scaling factor used in other works \cite{Borschevsky:13}.  


\begin{figure}[htb]
\includegraphics[width=0.45\textwidth]{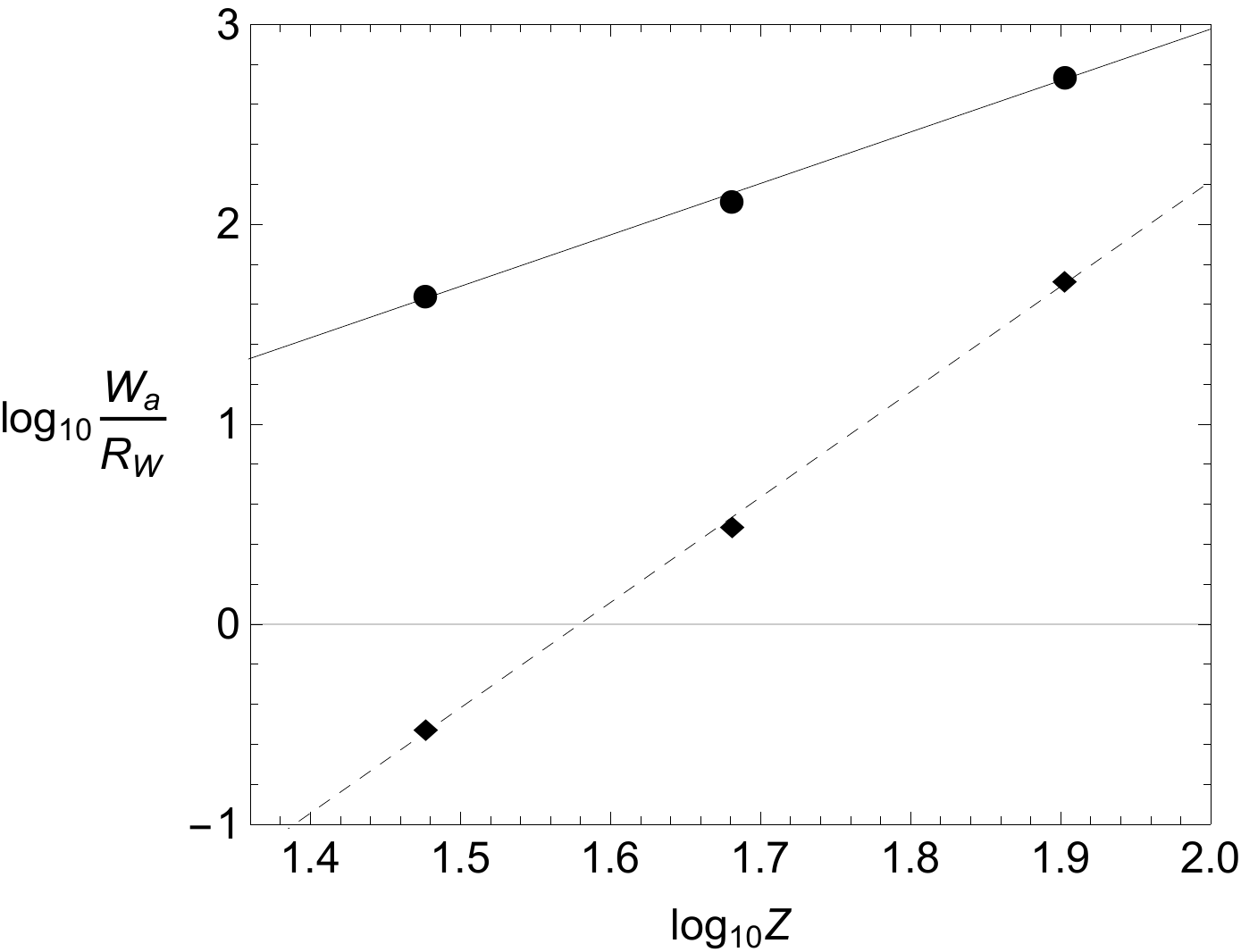}
\caption{\label{fig:logWa} Ratio of weak interaction constant $W_a$ to relativistic factor $R_W$ plotted against $Z$ for $Z = 30$, 48, and 80. Circles, solid line of best fit: ground state $^2\Sigma_{1/2}$; diamonds, dashed line of best fit: excited state $^2\Pi_{1/2}$. Calculated using the DHF method (see text).} 
\end{figure}

The ratio $W_a (^2\Sigma_{1/2})/R_W$ should scale linearly with $Z^{2}$, where $R_{W}$ is the relativistic factor defined by (\ref{eq:rel}). However, we observe (see Figure~\ref{fig:logWa}) a gradient of 2.5 instead of the expected gradient of 2. Similarly, we find $W_a(^2\Pi_{1/2})/R_W\sim Z^{5.3}$ rather than the expected $Z^4$. Both cases can be 
understood by the filling of the Hg atomic $d$ orbital close to the nucleus. Upon filling, the $d$ orbital expands relativistically, hence increasing effective nuclear charge and enhancing relativistic and NSD PNC effects. A similar trend is also seen in the $W_a (^2\Sigma_{1/2})$ constants for HgF, ZnF and CdF in \cite{Borschevsky:13}.

Furthermore, $W_a$ constants for the HgH $^2\Sigma_{1/2}$ and $^2\Pi_{1/2}$ electronic states have been calculated within the relativistic Fock-Space coupled cluster with single and double cluster amplitudes method. 35 outer-core and valence electrons were included in correlation treatment. For Hg and H atoms Dyall's uncontracted core-valence triple zeta (cv3z) basis sets \cite{Dyall:07,Dyall:12} were used. The $W_a$ constants were calculated at the equilibrium internuclear distance for the corresponding electronic states and are presented in the second column of Table~\ref{table:Wa}. There is good agreement between the DHF and CC methods of calculating $W_a$ constants in HgH with the two methods varying by 14\% for $W_a(^2\Sigma_{1/2})$ and 13\% for $W_a(^2\Pi_{1/2})$. 

\begin{table}[htb]
\caption{\label{table:Wa}Values for the effective weak interaction coefficients $W_a(^2\Sigma_{1/2})$ (ground state) and $W_a(^2\Pi_{1/2})$ (first excited state) calculated for group 12 hydrides. The $W_a(^2\Sigma_{1/2})$ for HgH is in good agreement with the semi-empirical estimates presented in \cite{Kozlov1985}}
\begin{ruledtabular}
\begin{tabular}
{p{1cm} p{1.7cm} p{1.7cm} p{1.7cm} p{1.7cm}}
Mol. &   DHF (Hz) & & CC (Hz) &  \\
         & $W_a(^2\Sigma_{1/2})$  & $W_a(^2\Pi_{1/2})$ & $W_a(^2\Sigma_{1/2})$& $W_a(^2\Pi_{1/2})$ \\
\hline
ZnH      & 61 &   -0.42 &- &-  \\
CdH     & 284 & -6.71 &- &-  \\
HgH      & 3882 & -372 & 3335 &-419 \\
\end{tabular}
\end{ruledtabular}
\end{table}



\section{PNC E1 amplitude}
The parity violating dipole transition amplitude $E1_{PNC}$ can be expressed as 
\begin{multline}
\label{amp}
\bra{i} E1_{PNC} \ket{k} = \\
\sum_{j} \frac{\bra{i} \vect{d}\cdot\vect{E}_0 \ket{j}\bra{j} H_{\rm eff} \ket{k}}{E_{k}-E_{j}} +
\frac{\bra{i} H_{\rm eff}\ket{j} \bra{j} \vect{d}\cdot\vect{E}_0 \ket{k}}{E_{i}-E_{j}},
\end{multline}
where $\vect{E}_0$ is the external electric field and $\vect{d}$ is the dipole moment operator.
In the first term of (\ref{amp}) \ket{j} and \ket{k} are sublevels of opposite parity situated in the ground $^2\Sigma_{1/2}$ electronic state.
In the second term of (\ref{amp}) \ket{i} and \ket{j} are sublevels of opposite parity corresponding to the $^2\Pi_{1/2}$ electronic state.
Expressions that appear in the decoupling of Eq. (\ref{amp}) can be found in the appendix of Ref.~\cite{Kozlov:91}, e.g.
for the present case ($I_1=I_2=1/2$)
\begin{multline}
\bra{JpF_1 F M} H_{\rm eff}\ket{J^\prime(-p)F_1^\prime F^\prime M} =\\
-\frac{1}{4}iW_a \kappa\,\delta_{F_1F_1^\prime}\delta_{FF^\prime}
\sqrt{\frac{3}{2}}
    \left\{
    \begin{array}{ccc}
    J &  J^\prime &  1 \\
    1/2 &  1/2 & F_1
    \end{array}
    \right\}\\
.(-1)^{F_1+J^\prime+1/2}\chi_JX_{JJ'},    
\end{multline}
where
\begin{gather}
    \chi_J= \pm p\,(-1)^{J-1/2},\nonumber \\
    X_{JJ}=(2J+1)\sqrt{\frac{2J+1}{J(J+1)}},\nonumber\\
    X_{JJ-1}=X_{J-1J}=\sqrt{\frac{(2J+1)(2J-1)}{J}}; \nonumber
\end{gather}
the plus sign is taken for the $^2\Pi_{1/2}$ state and the minus sign is taken for the $^2\Sigma_{1/2}$ state according to \cite{Kozlov:91}.

To obtain E1 and M1 amplitudes between $^2\Sigma_{1/2}$ and $^2\Pi_{1/2}$ electronic states we require the following matrix elements which we have calculated:
\begin{align}
 \label{Dperp}
D_{+} &= \bra{\Psi_{^2\Pi_{1/2}}}\vect{d}_{+} \ket{\Psi_{^2\Sigma_{-1/2}}} = 0.7~{\rm a.u.} \\
 \label{Gperp}
G_{+} &= \bra{\Psi_{^2\Pi_{1/2}}} \vect{L}_{+} - g_{S} \vect{S}_{+} \ket{\Psi_{^2\Sigma_{-1/2}}} = 1.4~{\rm a.u.}
\end{align}
Here $\vect{d}_{+}=\vect{d}_{\xi} + i\vect{d}_{\eta}$ is the dipole moment operator, $\vect{L}$ and $\vect{S}$ are the electronic orbital angular momentum and spin operators, and $g_{S} = -2.0023$ is the free-electron $g$-factor. Corresponding parallel components are small due to electronic configuration and are neglected here.

Matrix elements (\ref{Dperp}) and (\ref{Gperp}) have been calculated using the relativistic linear-response coupled cluster with single and double cluster amplitudes method \cite{Kallay:5} within the Dirac-Coulomb Hamiltonian. These calculations were performed at the internuclear distance which is the average of the $^2\Sigma_{1/2}$ and $^2\Pi_{1/2}$ equilibrium distances ($R=3.14$ Bohr \cite{Dufayard:88,Mosyagin:01b}). 


For the correlation calculation we used the {\sc mrcc} \cite{MRCC2013} and {\sc dirac15} \cite{DIRAC15} codes. For calculation of matrix elements over molecular bispinors the code developed in Refs.~\cite{Skripnikov:15b,Skripnikov:16b,Petrov:17b} was used.

\section{Results and discussion}
We have chosen to study a transition that occurs between the zeroth vibrational levels of the HgH molecule; this is because the ($\nu_X=0\to\nu_{A_1}=0$) transition has the maximal value of the square of vibration wave functions overlap (Frank-Condon factor) which is 0.5 \cite{Nedelec:88} and should result in a stronger transition compared to other vibrational states.

To calculate the circular polarization parameter $P=2\,\textrm{Im}(E1_{PNC})/{M1}$ we consider the ground rotational levels in both electronic states, set $F_1=0$ and use the following estimates for the energy separation between levels of opposite parity, $\Delta E$:
\begin{gather*}
\Delta E(^2\Sigma_{1/2})=2B(^2\Sigma_{1/2})-\gamma=8.64~\textrm{cm}^{-1},\\ 
\Delta E(^2\Pi_{1/2})=\Delta=3.36~\textrm{cm}^{-1},
\end{gather*}
where the experimental constants $B(^2\Sigma_{1/2})= 5.3888~{\rm cm}^{-1}$, $\gamma=2.14~{\rm cm}^{-1}$ and $\Delta=3.36~{\rm cm}^{-1}$ were taken from Ref.~\cite{Huber:79}.

It should be noted that the hyperfine splitting is considerably smaller than the rotational constant $B$ for both electronic states under consideration, e.g.~$A_{1, ||}(^2\Sigma_{1/2})$ is about 20 times smaller than the rotational constant for the $^2\Sigma_{1/2}$ state. Therefore, we neglect it below. For more accurate estimates one should numerically diagonalize the spin-rotational Hamiltonian (\ref{SROT}).

Furthermore, the estimated uncertainty of the calculated $W_a$ parameters are 15-20\%. This can be minimized considerably by applying combined technique developed in Refs.\cite{Skripnikov:16b,Skripnikov:17a,Skripnikov:17b}, but for our current purposes it is enough.

Using the aforementioned energy separations, the matrix elements (\ref{Gperp}, \ref{Dperp}), coupled-cluster $W_a$ constants for HgH from Table~\ref{table:Wa}, and neglecting possible phase difference in the terms in Eq.~(\ref{amp}) we obtain our final result
\begin{equation}
P=3 \cdot 10^{-6} \kappa.
\end{equation}
The leading contribution comes from the mixing of opposite parity levels of $^2\Sigma_{1/2}$ state, which is about 3 times larger than the term due to the mixing of opposite parity levels of $^2\Pi_{1/2}$ state.

\section{Conclusion}
The $^{199}$HgH molecule is a good candidate for PNC optical rotation experiments as it has closely spaced levels of opposite parity as well as a rotational constant large enough to resolve optical transitions between those levels. The circular polarization parameter was calculated to be $P=3 \cdot 10^{-6} \kappa $ which is 2 to 3 orders of magnitude larger than the estimated value for NSD PNC effects in atomic Xe, Hg, Tl, Pb and Bi \cite{Khriplovich:91,Dzuba2012}. 
Furthermore, HgH gives a pure NSD PNC signal needing a single transition for measurement; 
in contrast atomic experiments also give a much larger NSI PNC signal, requiring measurements on at least two different hyperfine transitions to isolate the small NSD PNC effect, which increases noise and possibly systematic effects. 

\section{Acknowledgments}
L.S. is grateful to Saint-Petersburg State University for a travel grant 11.42.700.2017 and RFBR, according to the research project No.~16-32-60013 mol\_a\_dk. A. G. is grateful for the support of the Australian Government Research Training Program Scholarship. L.S. and A.B. acknowledge the support of the Gordon Godfrey Visiting Fellowship. V.F. is grateful to the Australian Research Council for support. 


\end{document}